# A Multifunctional Array System Based on Adjustable-Phase Antenna for Wireless Communications

Guangwei Yang, *Member, IEEE*, Qiao Cheng, *Member, IEEE*, Jianying Li, *Member, IEEE*, Shuai Zhang, *Senior Member*, Steven Gao, *Fellow IEEE*, and Xiaodong Chen, *Fellow IEEE*

*Abstract*—In this work, an innovative method for controlling the current distribution of the radiating patch by adjusting the input phase is investigated to achieve both pattern and polarization reconfigurable characteristics for the multifunction. A compact and low-profile antenna with four fed ports is designed to implement the proposed method, which can operate linear, right-hand circular polarization (RHCP) and left-hand circular polarization (LHCP) with different beam directions in the operating band from 4.0 to 5.0 GHz. Even more, a four-by-four passive planar array is designed and fabricated based on this antenna element, which can scan the coverage of ±70° with low gain fluctuation and low sidelobe with dual-polarization. Meanwhile, it can realize the wide-angle scanning capability up to ±60° with low sidelobe with RHCP and LHCP. More important, the dual- and triple-beam with different directions can be obtained by the proposed array. Good agreement has been shown between measured and simulated results. Therefore, the proposed antenna is a good solution for wireless communication systems because of its simple-configuration, multifunction, and beamforming capability.

## I. Introduction

Due to the rapid and successive technological development of wireless communication systems, the antennas, as a key component for modern wireless communications, need to meet higher and higher performance requirements to satisfy complex communication environments such as multi-path effects, polarization mismatch, etc. Hence, designing multifunctional and intelligent array antenna is important. As one of the solutions, reconfigurable antennas have been used to achieve this goal because of the flexibility and versatility.

Reconfigurable antennas have been the subject of much attention and research in recent years, and fruitful results have been achieved. Reconfigurable antennas are mainly divided into four types: pattern [1], polarization [2], frequency [3], and hybrid (including polarization and pattern [4], polarization and frequency [5], pattern and frequency [6], and all three [7]) reconfigurable characteristics. The above reconfigurable antennas are typically implemented using electronic switches such as diodes, capacitors, etc., and mechanical operations like liquid metal, rotating structures, filling water, etc. However, there are some design bottlenecks for some reconfigurable antennas, including compactness, simplicity, and efficiency. Additionally, most reconfigurable antennas are not suitable for array applications.

For array antennas, pattern reconfigurable technology is applied to enhance beam steering capability or multi-beam to improve the signal-to-noise ratio of the system. A microstrip Yagi reconfigurable antenna element is designed and used in the array for broadening its scanning range [8]. Based on the O-slot patch element with p-i-n diodes, a 4-by-4 array with multi-beam is designed in [9]. A reconfigurable one-to-four power divider is used into an array for the multi-beam with the CP [10]. Besides, other pattern-reconfigurable technologies such as the parasitic antenna element with p-i-n diodes [11], radio frequency microelectromechanical systems switch the phase shifter [12], and digital coding characteristics [13] are used in the array to improve its performance. Also, frequency-reconfigurable technologies such as the frequency-reconfigurable slot-ring antenna with 16 p-i-n diodes [14] and a compact reconfigurable antenna element [15] are implemented in the array to realize beam-steering capability with dual-band. These designs enhance the range of applications for array antennas by implementing reconfigurable characteristics in the array, but the intricacy of the design and the array's radiation efficiency are both impacted by the rise of active devices. In [16], combining a dielectric resonator antenna (DRA) and a patch antenna, a phase-controlled pattern reconfigurable antenna is designed and used to realize wide-angle scanning capability, which realizes a passive pattern-reconfigurable characteristic without any active device. Additionally, reconfigurable radiating modes are used to change the phase center of the antenna for adjusting the pattern of the array [17][18]. The four-arm curl antenna is used to a phased array with high-tilt switching beam capability [19]. A multiple mode dipole array based on four elements is designed to realize beam steering capability in [20]. However, these designs either sacrifice some other performance such as the operating bandwidth, as well as the profile or some application defects. The dual-port phase antenna [21] is presented to realize multi-beam capability for improving the scanning capability. However, when facing complex communication scenarios, the above designs are difficult to meet requirements.

To confront these issues, in this work, an approach based on adjusting the input phase is investigated to achieve both pattern and polarization reconfigurable characteristics and applied to design a multifunctional array antenna. The highlights of this work include:

*a)* A wide-band (22.2%) and low-profile (0.11λ) antenna is developed to realize to two reconfigurable characteristics using a new passive technique, which differs from the above works [1]-[7], not only with high efficiency but also for array applications.

*b)* The antenna can realize pattern and polarization reconfigurable characteristics: *i)* it realizes the pattern reconfigurable capability including the wide-beam, dual-beam, ±50° beams with wide and narrow beam-width for each linear polarization; *ii)* the pattern can be directed in different directions with RHCP and LHCP, respectively.

*c)* For the array application, the proposed work can achieve polarization reconfigurable characteristics including dual linear polarization, RHCP, LHCP with wide-angle scanning capability. In the dual-polarization state, the array can achieve a scanning coverage of ±70°. Also, the pattern reconfigurability extends the wide-angle scanning capability and improves the sidelobe of the scanning beam.

*d)* The multi-beam capability of the array based on the proposed method is also obtained, which can be symmetrical/asymmetrical dual-beam, triple-beam etc.

Manuscript received August 2022. This work has been supported by the UK Royal Society Newton International Fellowship under Grant NIF/R1/191365.

Guangwei Yang and Xiaodong Chen are with the School of Electronic Engineering and Computer Science, Queen Mary University of London, London E1 4NS, UK. (gwyang086@gmail.com)

Jianying Li are the School of Electrical and Information, Northwestern Polytechnical University (NWPU), Xi'an, 710129, China.

Qiao Cheng is with the Department of Electronic and Electrical Engineering, College of Engineering, Design and Physical Sciences, Brunel University London, UB8 3PH Uxbridge, UK.

Shuai Zhang is with the Department of Electronic Systems, Aalborg University, Denmark.

Steven Gao is with the Department of Electronic Engineering, the Chinese University of Hong Kong, China.

## II. Guidelines For Antenna Design and Principle

### A. Principle and antenna design

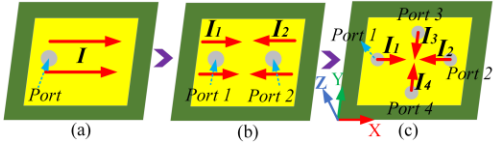

Fig. 1. Diagram of the design principle: (a) Single port; (b) Dual ports; and (c) Four ports.

To briefly explain the design mechanism, as shown in Fig. 1(a), the current density on the traditional patch is ***J***, which a classic radiating field of the patch antenna. For controlling the current distribution of the patch, two ports are used in the antenna for the two excited currents on the patch, as shown in Fig. 1 (b), which can control the current distribution of the antenna through the different input phases. Hence, the current density of the patch is given as

$$J_L = J_1 + J_2 \quad (1)$$

If the current density of every port in the proposed antenna has the same amplitude and different phase, so it can be

$$J_L = J_1 + J_2 = J(1 - e^{-j\Delta}) \quad (2)$$

Where $\Delta$ is the phase difference between two input ports of the antenna. Therefore, the proposed radiation field is the function of the input phase difference ($\Delta$). And the radiation pattern is written as: [22]

$$F_L = F_L(\theta, \Delta) \quad (3)$$

Which can be changed by the input phase difference ($\Delta$). Also, the proposed method can be used to the circular polarization antenna, as shown in Fig. 1(c), four ports are set to excite the patch antenna. To realize the circular polarization pattern, the phase difference between port 1(2) and port 3(4) should be $\beta$ ($\beta$=around ±90°). Hence, the current density of every port is $J_i$ ($i$=1, 2, 3 and 4) and the current of the antenna is given as

$$J_C = J_1 + J_2 + J_3 + J_4 = J_x(1 - e^{-j\Delta}) + J_y e^{-j\beta}(1 - e^{-j\Delta}) \quad (4)$$

Hence, the CP pattern is

$$F_C = F_C(\theta, \beta, \Delta) \quad (5)$$

And the polarization state is up to $\beta$. Hence, the pattern reconfigurable characteristic of the CP antenna can be realized by this method. More importantly, this method can also be applied to array antennas to improve beam scanning performance for polarization reconfigurable array and multi-beam radiating capability.

However, to implement the above mechanism, a high-quality antenna should be designed, which should not only have four ports, but good isolation among each port. More importantly, the wide beam capability should be considered. As shown in Fig. 2, the proposed antenna design process is presented. The antenna is developed from a patch antenna which has wide beam capability [22] but poor impedance matching. So a coupling feeding way is applied in Fig. 2 (b). In Fig. 2 (c), two-port way is designed, but the antenna's impedance deteriorates due to the addition of the port as shown in Fig. 3. Then a slot is used to improve the impedance matching as shown in Fig. 2 (d). Therefore, the four-port patch antenna is designed and optimized as shown in Fig. 2 (e).

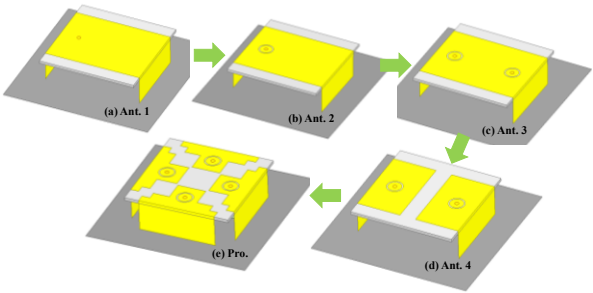

Fig. 2. The antenna design evolution process.

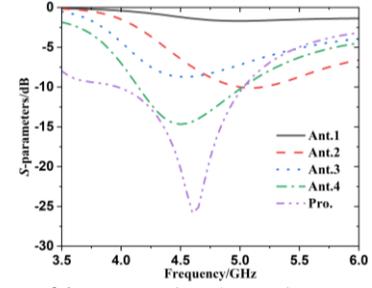

Fig. 3. S-parameters of the antennas (Ant.1 to pro.).

### B. Pattern-reconfigurable dual-polarization antenna

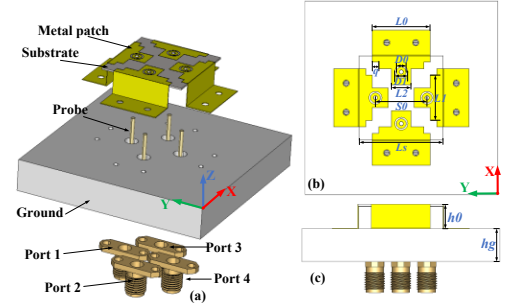

Fig. 4. Geometry of the proposed dual-polarization antenna: (a) 3D exploded view; (b) Plan view; (c) Side view; L0=18, L1=14, L2=6, Ls=26, S0=16, D0=1.5, D1=2, q=2, h0=7.5, (unit: mm).

TABLE I
DUAL-POLARIZATION RECONFIGURABLE ANTENNA OPERATION

| No. | $\Delta$ | Pattern model | HPBW |
|---|---|---|---|
| a | 0° | Dual-beam | / |
| b | 45° | Direct to -50° with narrow beam | 85° |
| c | 90° | Direct to -50° | 96° |
| d | 135° | Direct to -45° with wide beam | 94° |
| e | 180° | Wide beam | 228° |
| f | 225° | Direct to 45° with wide beam | 95° |
| g | 270° | Direct to 50° | 96° |
| h | 315° | Direct to 50° with narrow beam | 86° |

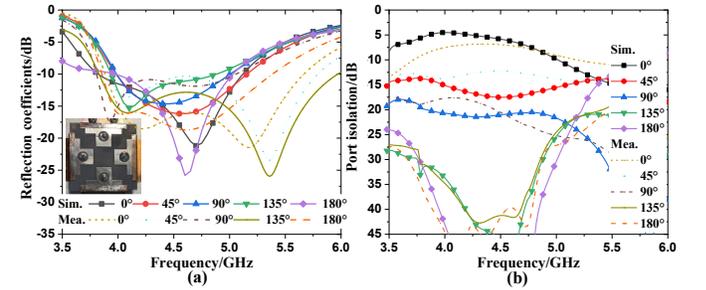

Fig. 5. *S*-parameters of the proposed dual-polarization pattern-reconfigurable antenna: (a) Reflection coefficients; and (b) Port isolation.

As shown in Fig. 4. The proposed antenna comprised of four special bended metallic plates (SBMP), a substrate with the thickness of 0.762 mm and permittivity ($\varepsilon$) of 2.2, and the ground plane. One part of the SBMP is printed on the substrate and the other part is bended to connect to the ground plane. As shown in Fig. 4 (a), four SBMPs are fed by four copper probes, respectively. The key design parameters and stack diagram of the proposed reconfigurable antenna are given in Fig. 4 (b) and (c). Based on the above principle, the feeding-mechanism of the proposed dual-polarization antenna is depicted. The input power is equally divided by a power divider and feeds into two separate ports (Port 1 and Port 2) of the proposed antenna. A phase shifter is added to Port 2, resulting in phase difference ($\Delta$) between the two ports. Other ports (Port 3 and Port 4) connect to matching loads. Combined with Eq. (2), note that the phase difference $\Delta$ decides the phase distribution of the antenna, which radiates different radiation patterns. Hence, the relationship between the pattern-reconfigurable modes and the phase difference is shown in Table I. The phase variation ranges from 0 to 360°. Similarly, the pattern reconfigurable characteristic is also achieved by the proposed antenna with another polarization in another plane.

As given in Fig. 5, the operating frequency band of the proposed dual-polarization antenna is given. Note that there are still some differences in the frequency bands of the different operating modes as the input phase changed. However, their same operating band is from 4.0 to 5.15 GHz (25.1%). The measurements can cover this operating band and are wider than the simulated frequency band due to the interference in the feeding network causing a wider resonant impedance. Besides, the phase shifter and power divider are used to measure, and the transmitted line for the phase difference and power division network are used to simulate. The simulated and measured port isolations are also drawn in Fig. 5. Noted that the port isolation of the antenna is affected by the phase difference between the two ports, which is worse when the two ports are in phase and gradually improves as the phase difference increases. To explore how this method is implemented by the antenna, the current distributions with different phase differences (Δ) are depicted in Fig. 6. It is obviously observed that the antenna has different current distribution since the different input phase difference between two ports. From Fig. 6 (a) to (d), it can radiate a monopole-like pattern, a wide beam pattern, and different directional patterns. Therefore, the current distribution of the proposed radiating structure is controlled by the input phase of the two excited ports, thus enabling reconfigurable radiation performance. In addition, the input amplitude of each port also affects the radiating characteristics, but this has a slight effect and can worsen cross-polarization if the port input is severely unequal.

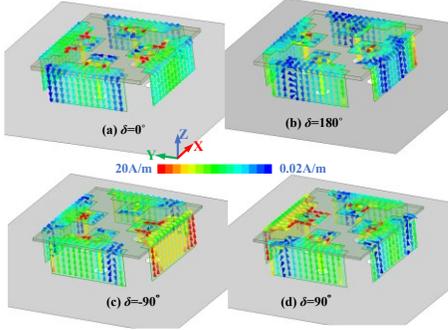

Fig. 6. Currents distributions of the dual-polarization pattern-reconfigurable antenna at 4.5 GHz: (a) Δ=0°; (b) Δ=180°; (c) Δ=-90° (270°); (d) Δ=90°.

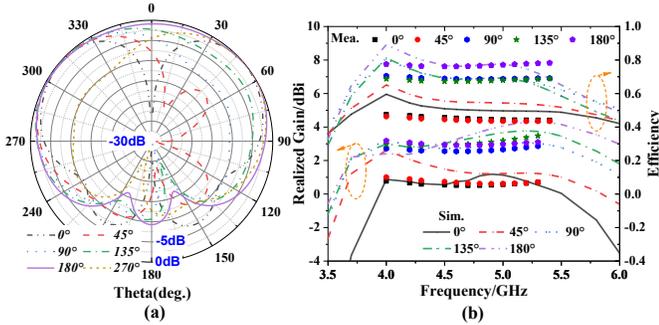

Fig. 7. (a) Normalized radiation patterns and (b) Realized gain and efficiency of the proposed dual-polarization pattern-reconfigurable antenna.

Fig. 7(a) shows the radiation patterns of the dual-polarization antenna. Based on the different phase differences, the antenna's pattern could be directed in different directions and have different beam-widths. The detailed reconfigurable modes with varying phase differences are given in Table I. Similarly, in *xoz*-plane, the proposed antenna can also implement the same pattern-reconfigure modes on another polarization. The measured radiation patterns with varying phase differences are shown in Fig. 8. Noted that the simulation results agree very well with the test results. In addition, the cross-polarization of the patterns is below -10 dB, especially in the wide beam pattern, which is below -30 dB. The realized gain and efficiency of the proposed dual-polarization antenna with different input phases are shown in Fig. 7 (b). Note that the gain is not very high due to the wide beam, which is around 3 dBi. And the gain of the antenna decreases when the beam directs to large angles. Meanwhile, the efficiency is around 75% in the operating frequency band and drops when the beam directs to large angles.

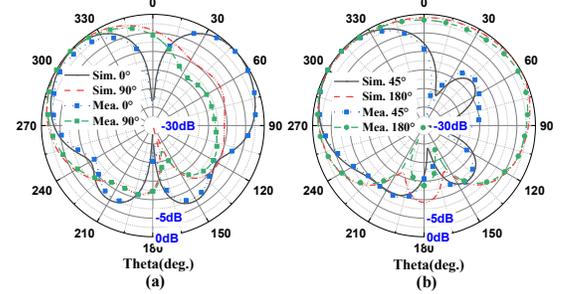

Fig. 8. Comparing the simulated and measured normalized radiation patterns of the proposed dual-polarization pattern-reconfigurable antenna.

### C. Reconfigurable circular polarization antenna

Similarly, the pattern-reconfigurable CP antenna is realized by exciting four ports based on the design theory. The input power is equally divided by a power divider and feeds into four separate ports of the proposed antenna. The phase shifters are added to Port 2, 3, and 4, resulting in phase differences among the four ports. To ensure the antenna's proposed circular polarization characteristics, the phase difference (β) is around ±90°, and the phase difference (Δ) is adjustable to realize the pattern reconfiguration which is shown in Table II. The *S*-parameters of the proposed CP antenna at different modes are reported in Fig. 9, which can still operate in the bandwidth from 4 to 5.0 GHz, both in polarization- and pattern-reconfigurable modes. Since the LHCP characteristics are similar to the ones of RHCP as seen in Fig. 9 (a), The RHCP axial ratio (AR) of the antenna is only given as shown in Fig. 9 (b) for brevity. It can be found that in the proposed operating band, the axial ratio of the antenna satisfies the requirements of the CP antenna.

TABLE II
CIRCULAR POLARIZATION RECONFIGURABLE ANTENNA OPERATION

| β | Δ | Pattern model | HPBW/AR-HPBW |
|---|---|---|---|
| 90° | 0° | Broadside wide beam of RHCP | 155°/114° |
|  | 45° | Direct to -30° of RHCP (R.-M.) | 138°/110° |
|  | -45° | Direct to 30° of RHCP(L.-M.) | 130°/102° |
| -90° | 0° | Broadside wide beam of LHCP | 156°/116° |
|  | 45° | Direct to -30° of LHCP (R.-M.) | 138°/108° |
|  | -45° | Direct to 30° of LHCP(L.-M.) | 134°/102° |

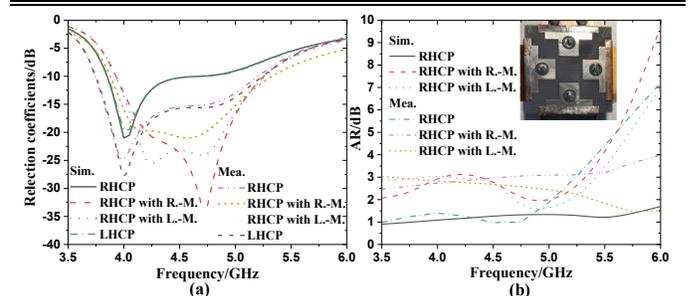

Fig. 9. Measured and simulated (a) Reflection coefficients and (b) AR of the proposed circular polarization pattern-reconfigurable antenna.

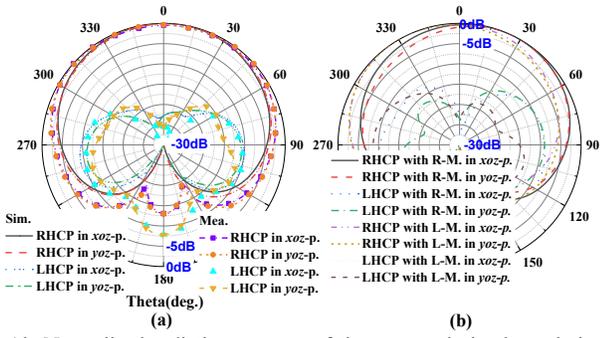

Fig. 10. Normalized radiation patterns of the proposed circular polarization antenna with RHCP at 4.5 GHz: (a) Radiation patterns; (b) Different modes.

As shown in Fig. 10 (a), the radiation pattern of the proposed antenna with the RHCP are reported. As obtained from the current distribution, this antenna can radiate a wide beam pattern with the 3-dB beam-width of 155° and the AR pattern with the 3-dB-beam-width is also wide and up to 118°. Besides, the cross-polarization is very low. Besides, the measured results are also given. The measured results agree with the simulations. The characteristics of the wide beam and low cross-polarization have been verified. The measured realized gain is about 2.2 dBi in the operating frequency band, and the efficiency is about 75%. The different reconfigurable patterns based on Table II are shown in Fig. 10 (b), which can be directed to around ±30°. Despite the radiation patterns pointing to the other direction, its cross-polarization is unaffected and remains relatively good. Similarly, the good radiation characteristics with LHCP are also realized.

## III. Planar Array

As described in Section II-A, the proposed method can be used to array antenna to realize polarization-reconfigurable wide-scanning capability and multi-beam patterns. Hence, a four-by-four planar array is implemented for these applications based on the above antenna element. Thanks to the compact element size, the inter-element distance of the array is 30 mm ($0.5\lambda_0$ at 5.0 GHz). To explore the proposed array's polarization-reconfigurable beam scanning capability, a simple analysis with this planar array is presented in this section. As all we know, the ideal array field pattern is calculated as

$$F_{total}(\theta) = \sum_{m=0}^{M-1}\sum_{n=0}^{N-1} F_{mn}(\theta) e^{-ja_{mn}} e^{jk(md_x\cos\varphi + nd_y\sin\varphi)\sin\theta}$$
$$= \sum_{m=0}^{M-1}\sum_{n=0}^{N-1} F_{mn}(\theta) e^{-j(ma_x + na_y)} e^{jk(md_x\cos\varphi + nd_y\sin\varphi)\sin\theta} \quad (6)$$

Where the input phase difference of the array element is $a_{mn} = ma_x + na_y$, which, usually, could decide the beam direction. However, the scanning gain fluctuation is up to $F_{mn}(\theta)$. It can be equivalent to

$$F_{total}(\theta) = F(\theta) \cdot AF(\theta) \quad (7)$$

Where $AF(\theta, \varphi)$ is the planar array factor. Noted that the beam direction of the array is up to the phase difference in the array factor, but the array pattern can be optimized by the phase difference ($\Delta$) based on the different scanning direction. Meanwhile, the polarization of the array is decided by the phase $\beta$ ($\beta = \pm 90°$). Therefore, in this array, we combine the control of the radiation performance of the antenna element with the phased array to achieve the optimization of the array polarization-reconfigurable beam scanning capability.

### A. Dual-polarization planar array

Due to the limitations of our test conditions, two different feeding ways are available for each of the two polarization states. For demonstration, as shown in Fig. 11 (a), a four-by-four planar array is manufactured and connected to eight one-to-four feed networks, eight mechanical phase shifters, and a one-to-eight power driver. The passive planar array system is realized and tested in the anechoic chamber to verify the beam scanning capability. To show the array excited method clearly, eight different colored lines are used in Fig. 11, with the remaining ports connected to matching loads. Similarly, the scanning capability of the other polarization is measured. Fig. 12 shows the measured *S*-parameters of the proposed array at different scanning angles. The array operates with a bandwidth of 4 to 5.1 GHz over most scan angle ranges. But the impedance matching of the array becomes poor when scanning to around 70°. This is because the relatively large input phase leads to stronger induced currents between the array elements in the array, resulting in poorer impedance matching of the array elements. Overall, the reflection coefficients at large scanning angles up to 70° is lower than -6 dB in the proposed bandwidth and the ones at other scanning angles are lower than -10 dB. Besides, the post isolation and mutual coupling in the array are also presented, which are lower than -18 dB and -15 dB to guarantee the good scanning performance, respectively. To demonstrate the effect of the phase difference ($\Delta$) on the performance of the array radiation pattern, the array patterns with and without optimized phase difference ($\Delta$) when the beam is scanned to 60° are given in Fig. 12 (b). Comparing these patterns shows a significantly improved gain (up to 3 dB) and sidelobe levels (lower than -11 dB) for the proposed approach when the array is scanned at large angles. This is because that the scanning pattern can be optimized by changing the antenna element's radiation pattern in the array to achieve a better scanning pattern.

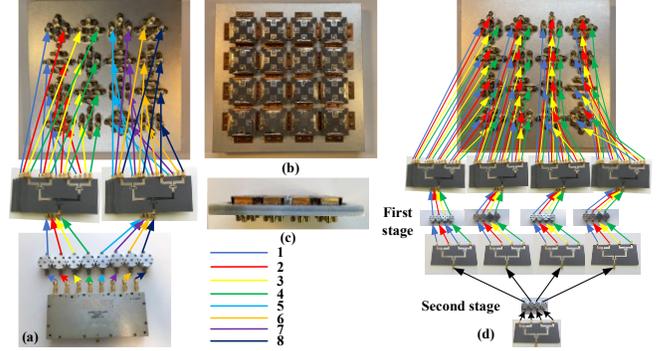

Fig. 11. Photograph of the proposed array and measurement: (a) Passive array system and linear polarization excitation; (b) Plan view of the array; (c) Side view of the array; (d) CP array excitation.

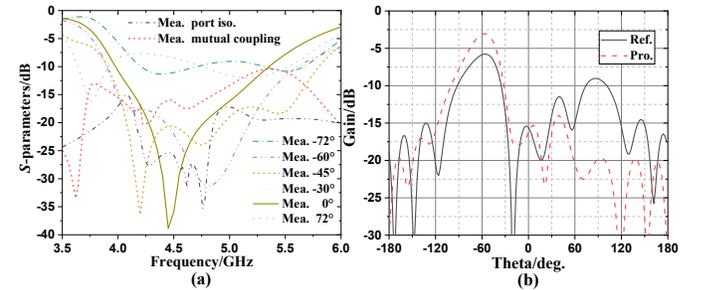

Fig. 12. (a) Measured *S*-parameters of the proposed dual-polarization antenna at different scanning angles and (b) Normalized radiation patterns with and without optimized phase difference ($\Delta$) at directing to -60°.

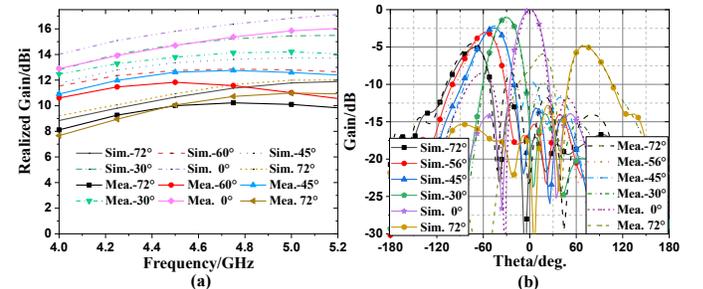

Fig. 13. Scanning performance of the array antenna: (a) Scanning capability in the bandwidth; (b) Scanning normalized pattern at 4.5 GHz.

The measured and simulated scanning characteristics are depicted in Fig. 13. Note that the proposed design can realize the scanning capability from -70° to 70° and the gain reduction is lower than 3 dB in the range from -60° to 60° in the proposed bandwidth. To verify the scanning performance, the pattern at the center frequency is just shown in Fig. 13 (b). It is found that the proposed array has good scanning capability with low sidelobe and wide-angle scanning range. The measured results are also given, which are in general agreement with the simulation results.

*B. Circular polarization planar array*

Based on the CP antenna element excitation way in Section in Section II-C, as shown in Fig. 11 (d), four one-to-four feeding networks are connected to a column of antenna elements in the array, meanwhile, which are connected to the first stage phase shifters, ensuring the CP characteristics and the deflection of the radiation pattern of the array elements, respectively. Hence, the passive CP array system is set up and measured in the anechoic chamber.

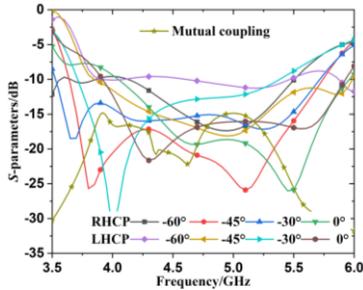

Fig. 14. Measured total *S*-parameters of the proposed CP array at different scanning angles and mutual coupling in the array.

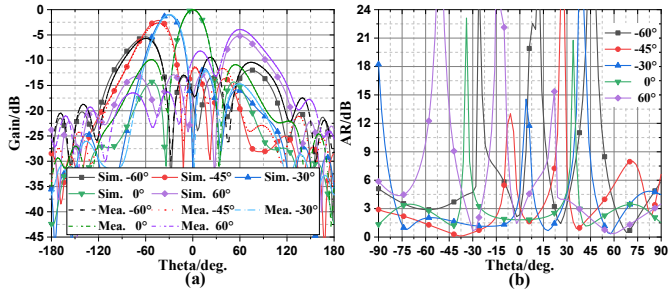

Fig. 15. (a) Measured and simulated normalized radiation patterns and (b) AR patterns of RHCP with scanning different angles.

In Fig. 14, the measured *S*-parameters of the proposed CP array at difference scanning angles are presented, which operates from 4 to 5.1 GHz in the scanning range and the mutual coupling between the adjacent units in the array is lower than -15 dB. We also give the *S*-parameters for the LHCP, showing that the impedance matching still works in the polarization reconfigurable mode. Similarly, the far field characteristics of the array are verified by the RHCP for the sake of brevity. The measurements and simulations about the radiation patterns and AR patterns of the RHCP are shown in Fig. 15. Note that the array can still achieve beam scanning capability within ±60° in its operating frequency band. The sidelobe level of the beam is less than -10 dB except for the sidelobe level below -7.5 dB at 60°. Besides, we can see that the test results are in general agreement with the simulated results. The AR of the beam is less than 3 dB regardless of the scanning angle, and in particular, the AR of the scanning beam is better when scanned to around -30°. This is because the optimization of the phase difference within the array element results in excellent AR performance in this direction. Again, as the pattern-reconfigurable performance is not very good when scanning to larger angles, the AR is worse than the beam at smaller angles but still less than 3 dB. In Fig. 16, the AR and the realized gain of the array in the frequency band are reported. Noted that the proposed array antenna has good far-field characteristics, where's AR is less than 3 dB in the frequency band and the beam gain gradually decreases with increasing the scanning angle.

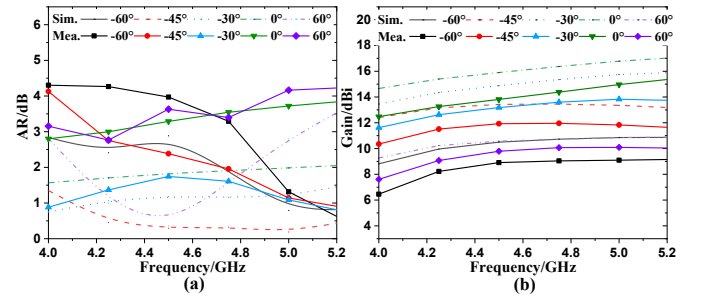

Fig. 16. Far field performance of the proposed array: (a) AR and (b) Realized gain of RHCP with frequency.

*C. Multi-beam capability*

To explore the proposed array's multi-beam capability, a simple analysis with this planar array is presented in this section. As all we know, the ideal array field pattern is calculated as shown in Eq. (6). Based on Eq. (5), we can find that the pattern of the antenna element could be tilted by manipulating the phase difference ($\Delta$), as shown in Table I, which has many models ($F_{mn}(\theta) \in (F_a, F_b \cdots F_h)$). Hence, to enhance the scanning capability of the array, based on the beam scanning direction, the pattern $F_{mn}(\theta)$ of the element can be modified by the phase difference ($\Delta$), which is a new dimension to improve the scanning capability. However, in an infinite array antenna, since the pattern of the array elements tends to be the same, so the proposed method will gradually weaken, and it is only suitable for small arrays.

In addition, no limited by the size of the array, $F_a(\theta)$ and $F_e(\theta)$ can be implemented in the array, so $F_{mn}(\theta) \in (F_a, F_e)$, and we can use the optimization algorithm [23] to optimize Eq. (6) to get the multi-beam radiation.

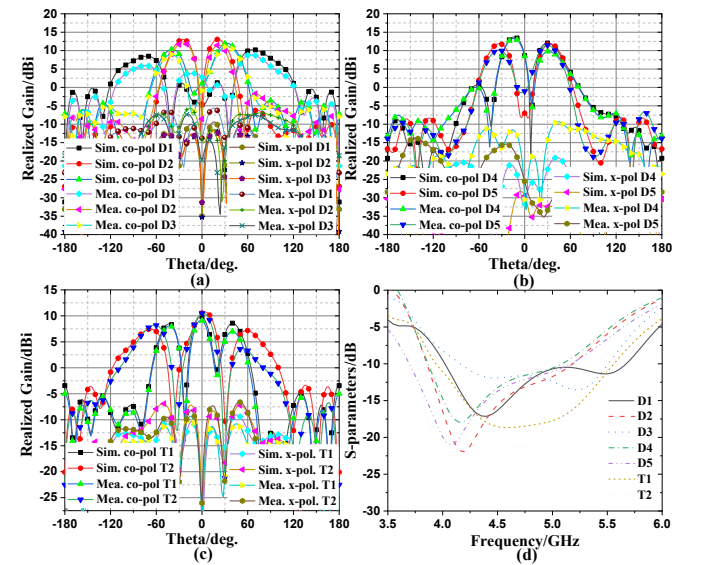

Fig. 17. Far field performance of the proposed array with multi-beam capability: (a) Dual-beam patterns (D1, 2, 3); (b) Dual-beam patterns (D4, 5); (c) Triple-beam patterns (T1, 2); (d) Measured *S*-parameters.

Then, the particle swarm optimization (PSO) [24] algorithm is employed to calculate the optimal solution. We can obtain the array element model or phase excitation coding matrixes for different beam radiation as shown in Table III. For the dual-beam, in Fig. 17 (a), the symmetrical dual-beam patterns are obtained by the proposed approach, which can get different array element model coding matrixes based on the different beam pointing. Also, the dual-beam pattern in H-plane and the asymmetrical dual-beam pattern are obtained as given in Fig. 17 (b). For the triple-beam, the patterns are shown in Fig. 17 (c). Different beam pointing of the patterns are up to the different array element model coding matrixes. Therefore, if the dimension of the proposed array is sufficiency large, arbitrary beams can be obtained by the proposed design. The measured results are also drawn in Fig. 17, noted that they are similar to the simulated results and the operation bandwidth is always from 4 to 5.1 GHz. The simulated 3D dual- and triple-beam patterns are shown in Fig. 18, noted that the proposed array achieves a good multi-beam capability.

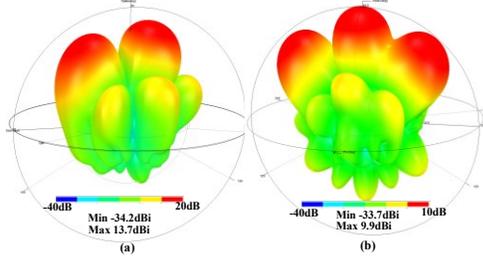

Fig. 18. 3D far field performance of the array with multi-beam capability: (a) Dual-beam asymmetrical pattern (D4); (b)Triple-beam pattern (T1).

TABLE III
MULTI-BEAM CHARACTERISTICS OF THE PROPOSED ARRAY

| Type | Beam pointing | Coding matrix |
|---|---|---|
| D1 | -70° and 70° | [$F_a$, -$F_a$, $F_a$, -$F_a$,] |
| D2 | -25° and 25° | [$F_a$, $F_a$, -$F_a$, -$F_a$,] |
| D3 | -34° and 34° | [$F_a$, -$F_a$, -$F_a$, $F_a$,] |
| D4 | -15° and 30° | [$F_a$, -$F_a$, -$F_a$, $F_a$,] with α of 150° |
| D5 | -30° and 30° | [$F_a$, -$F_a$, -$F_a$, $F_a$,]$^T$ |
| T1 | -45°, 0°, 45° | [-$F_a$, $F_e$, $F_e$, $F_a$,] |
| T2 | -65°, 0°, 65° | [$F_e$, -$F_a$, $F_e$, -$F_a$,] |

## IV. CONCLUSION

In this work, an innovative method based on controlling the current distribution of the radiating patch by adjusting the input phase has been presented to realize both pattern and polarization reconfigurable characteristics of the antenna and multifunctional array. A compact and low-profile antenna with four ports is designed to implement this method, which can operate at a wide bandwidth from 4.0 to 5.0 GHz to realize pattern and polarization reconfigurable characteristics. For the dual-polarization state, the antenna realizes the pattern reconfigurable capability including the wide-beam, dual-beam, ±50° beams for each polarization. For the CP state, the beam can direct to different directions with wide beam for the RHCP and LHCP, respectively. A four-by-four passive planar array can realize polarization-reconfigurable characteristic with beam scanning capability including the beam scanning over ±70° for the dual-polarization and over ±60° for the CP, respectively. In particular, the proposed method can be used to realize the multi-beam capability.

## REFERENCES

[1] G. Yang, J. Li, D. Wei, S. Zhou and R. Xu, "Pattern reconfigurable microstrip antenna with multidirectional beam for wireless communication," *IEEE Trans. Antennas Propag.*, vol. 67, no. 3, pp. 1910-1915, March 2019.
[2] Q. Chen, J. Li, G. Yang, B. Cao and Z. Zhang, "A polarization-reconfigurable high-gain microstrip antenna," *IEEE Trans. Antennas Propag.*, vol. 67, no. 5, pp. 3461-3466, May 2019.
[3] S. -C. Tang, X. -Y. Wang and J. -X. Chen, "Low-profile frequency-reconfigurable dielectric patch antenna and array based on new varactor-loading scheme," *IEEE Trans. Antenna Propag.*, early access.
[4] G. Yang, J. Li, B. Cao, D. Wei, S. Zhou and J. Deng, "A compact reconfigurable microstrip antenna with multidirectional beam and multipolarization," *IEEE Trans. Antenna Propag.*, vol. 67, no. 2, pp. 1358-1363, Feb. 2019.
[5] P. Qin, Y. J. Guo, Y. Cai, E. Dutkiewicz and C. Liang, "A reconfigurable antenna with frequency and polarization agility," *IEEE Antennas and Wireless Propagation Letters*, vol. 10, pp. 1373-1376, Dec. 2011.
[6] V. T. Bharambe, J. Ma, M. D. Dickey and J. J. Adams, "RESHAPE: a liquid metal-based reshapable aperture for compound frequency, pattern, and polarization reconfiguration," *IEEE Trans. Antennas Propag.*, vol. 69, no. 5, pp. 2581-2594, May 2021.
[7] L. Ge, Y. Li, J. Wang and C.-Y. D. Sim, "A low-profile reconfigurable cavity-backed slot antenna with frequency polarization and radiation pattern agility", *IEEE Trans. Antennas Propag.*, vol. 65, no. 5, pp. 2182-2189, May 2017.
[8] S. Xiao, C. Zheng, M. Li, J. Xiong and B. Wang, "Varactor-loaded pattern reconfigurable array for wide-angle scanning with low gain fluctuation," *IEEE Trans. Antennas Propag.*, vol. 63, no. 5, pp. 2364-2369, May 2015.
[9] M. Wang, S. Xu, N. Hu, W. Xie, F. Yang, Z. Chen, and M. Li, "Design and measurement of a ku-band pattern-reconfigurable array antenna using 16 o-slot patch elements with p-i-n diodes," *IEEE Antennas and Wireless Propagation Letters*, vol. 19, no. 12, pp. 2373-2377, Dec. 2020.
[10] M. Shirazi, J. Huang, T. Li and X. Gong, "A switchable-frequency slot-ring antenna element for designing a reconfigurable array," *IEEE Antennas and Wireless Propagation Letters*, vol. 17, no. 2, pp. 229-233, Feb. 2018.
[11] X. Cai, A. Wang, N. Ma and W. Leng, "A novel planar parasitic array antenna with reconfigurable azimuth pattern," *IEEE Antennas and Wireless Propagation Letters*, vol. 11, pp. 1186-1189, Sep. 2012.
[12] N. Kingsley, G. E. Ponchak and J. Papapolymerou, "Reconfigurable RF MEMS phased array antenna integrated within a liquid crystal polymer (LCP) system-on-package," *IEEE Trans. Antennas Propag.*, vol. 56, no. 1, pp. 108-118, Jan. 2008.
[13] X. G. Zhang, W. X. Jiang, H. W. Tian, Z. X. Wang, Q. Wang and T. J. Cui, "Pattern-Reconfigurable Planar Array Antenna Characterized by Digital Coding Method," *IEEE Trans. Antennas Propag.*, vol. 68, no. 2, pp. 1170-1175, Feb. 2020.
[14] M. Shirazi, J. Huang, T. Li and X. Gong, "A Switchable-Frequency Slot-Ring Antenna Element for Designing a Reconfigurable Array," *IEEE Antennas and Wireless Propagation Letters*, vol. 17, no. 2, pp. 229-233, Feb. 2018.
[15] N. Haider, A. G. Yarovoy and A. G. Roederer, "L/S-Band Frequency Reconfigurable Multiscale Phased Array Antenna with Wide Angle Scanning," *IEEE Trans. Antennas Propag.*, vol. 65, no. 9, pp. 4519-4528, Sept. 2017.
[16] Z. Chen, Z. Song, H. Liu, X. Liu, J. Yu and X. Chen, "A compact phase-controlled pattern-reconfigurable dielectric resonator antenna for passive wide-angle beam scanning," *IEEE Trans. Antennas Propag.*, vol. 69, no. 5, pp. 2981-2986, May 2021.
[17] S. Sharma, L. Shafai, B. Balaji, A. Damini, and G. Haslam, "Investigations on multimode microstrip patch antenna and phased arrays providing multiphase centres," in *Proc. IEEE APS-URSI*, vol. 1A, Washington, DC, USA, Jul. 2005.
[18] T. Mitha and M. Pour, "Principles of adaptive element spacing in linear array antennas," *Nature Sci. Rep.*, vol. 11, no. 1, pp. 1-11, 2021.
[19] H. Zhou et al., "Reconfigurable phased array antenna consisting of high-gain high-tilt circularly polarized four-arm curl elements for near horizon scanning satellite applications," *IEEE Antennas and Wireless Propagation Letters*, vol. 17, no. 12, pp. 2324-2328, Dec. 2018.
[20] N. R. Labadie, S. K. Sharma and G. M. Rebeiz, "A novel approach to beam steering using arrays composed of multiple unique radiating modes," *IEEE Trans. Antennas Propag.*, vol. 63, no. 7, pp. 2932-2945, July 2015.
[21] E. Liu, J. Geng, K. Wang, et al., "Generalized principle of pattern multiplication based on the phase antenna element," *2020 IEEE International Symposium on Antennas and Propagation and North American Radio Science Meeting*, 2020, pp. 353-354.
[22] G. Yang, J. Li, J. Yang and S. Zhou, "A wide beamwidth and wideband magnetoelectric dipole antenna," *IEEE Trans. Antennas Propag.*, vol. 66, no. 12, pp. 6724-6733, Dec. 2018.
[23] C. Feng, X. Li, Y. Zhang, X. Wang, L. Chang, F. Wang, et al, "RFlens: metasurface-enabled beamforming for IoT communication and sensing," In *Proceedings of the 27th Annual International Conference on Mobile Computing and Networking,* pp. 587-600, October 2021.
[24] D. W. Boeringer and D. H. Werner, "Particle swarm optimization versus genetic algorithms for phased array synthesis," *IEEE Trans. Antennas Propag.*, vol. 52, no. 3, pp. 771-779, March 2004.